\documentclass[aps,pre,preprint,superscriptaddress,graphicx]{revtex4-1} 

\usepackage{amsmath}
\usepackage{amssymb}
\usepackage{color}
\usepackage{graphicx}
\usepackage{rotating}

\begin{document}

\title{Confinement dynamics of a semiflexible chain inside nano-spheres}

\author{A. Fathizadeh }
\address{Sharif University of
Technology, Institue for nanoscience and nanotechnology, P.O. Box 14588-89694, Tehran,
Iran.}

\author{Maziar Heidari}
\address{School of Mathematics, Institute for Research in Fundamental Sciences (IPM), Tehran, Iran.}

\author{B. Eslami-Mossallam}
\address{Instituut-Lorentz for Theoretical Physics, P.O.Box 9506, 2300 RA Leiden, The Netherlands.}

\author{M.R. Ejtehadi}
\email{ejtehadi@sharif.edu}
\address{Sharif University of
Technology, Department of Physics, P.O. Box 11365-8639, Tehran,
Iran.}

\newpage

\begin{abstract}
We study the conformations of a semiflexible chain, confined in nano-scaled spherical cavities, under two distinct processes of confinement. Radial contraction and packaging are employed as two confining procedures. The former method is performed by gradually decreasing the diameter of a spherical shell which envelopes a confined chain. The latter procedure is carried out by injecting the chain inside a spherical shell through a hole on the shell surface. The chain is modeled with a rigid body molecular dynamics simulation and its parameters are adjusted to DNA base-pair elasticity. Directional order parameter is employed to analyze and compare the confined chain and the conformations of the chain for two different sizes of the spheres are studied in both procedures. It is shown that for the confined chains in the sphere sizes of our study, they appear in spiral or tennis-ball structures, and the tennis-ball structure is more likely to be observed in more compact confinements. Our results also show that the dynamical procedure of confinement and the rate of the confinement are influential parameters of the structure of the chain inside spherical cavities.  
\end{abstract}



\maketitle

\section{Introduction}

In the recent years conformation of the polymers inside confined structures has been investigated widely \cite{Marenduzzo03, Elrad01, Morrison01, Liu01, Cerda01,Ali02}. A category of these studies is polymer packaging in which a polymer chain is encapsulated inside a confined space. The examples of these highly confined structures are naturally found in viral capsids where a DNA strand is packed inside a space comparable to its persistence length. Thus, understanding the structure of a semiflexible chain in such severe confinements and the way the molecule responses to the geometrical constraints may help understand DNA function in viral capsids.

Many aspects of this phenomenon like the effect of packaging and ejection forces, packaging time, and shape of the confining space on the conformation of the DNA (single strand or double strand) are studied \cite{Micheletti01, Petrov02, Ali01, Marenduzzo01, Rollins01}. Simulations propose spool like conformation for a semiflexible polymer in the spherical cavity (See \cite{Petrov01,Petrov02,Lamarque01}). Such structures are also observed in the macroscopic scale in packing of elastic wires in spherical cavities \cite{Stoop01}. However, Spakowitz and Wang \cite{Spakowitz01} have reported toroidal conformations for a DNA in a spherical capsid by preventing the release of the twist at the end of the polymer chain.  

For polymers in spherical confinement, the bond vectors of the polymer, the vector along connection of two neighboring beads, tends to be tangent to the surface of the sphere. Then it is proposed that the different configurations of the confined polymers can be understood in an analogy with the nematic order of liquid crystal droplets with planar anchoring~\cite{Katzav01}.

It is well known that there is no defect-free structure of tangent lines on the surface of a sphere. According to the Poincare-Hopf theorem, the total charge of the defects is $+2$ \cite{Poincare01,Hopf01}. There are two possibilities for bipolar defect arrangements with  two $+1$ defects (spool or polar) and one tetrahedral of four $+1/2$ defects (Tennis ball) \cite{DNelson01,Mozaffari01}. These structures have been reported for confined polymers on the surface of a sphere \cite{Angelescu01, Zhang01}, except the polar structure which is not favorable, because of the excluded volume effects. Using Monte Carlo simulations, Zhang and Chen have shown that a self avoiding worm-like polymer tends to create tennis ball shapes on a spherical surface~\cite{Zhang01}. In a similar Monte Carlo study, Angelescu {\it et al.} \cite{Angelescu01} show that in the absence of long range electrostatic repulsion, the polymer shows spiral structure on the surface of the sphere, but by applying the long range repulsion, tennis-ball textures become more likely. Recently, Oskolkov {\it et al.} \cite{Oskolkov01} applied a so-called density functional theory based model and by using nematic ordering concept for a polymer confined inside a sphere, they predicted a spool structure for stiff and long confined polymer chains.

Our aim is to study the structure of a confined semiflexible polymer chain with equivalent elastic properties of a double stranded DNA in nano-scale spherical cavities. \textcolor{black}{We use a model which  is parameterized at base-pair level and size of a DNA base-pair is the characteristic length of the model.} The observed structures of the confined chain are demonstrated and studied with the nematic order analysis on the surface of the sphere. In addition to the effect of size of the cavity, we mainly focus on the dynamics of the confining procedure and we show that different dynamical procedures may lead to different structures for the confined polymer chain.

\section{Materials and Method}
\subsection{Interaction Potentials}

The polymer chain is built based on a coarse grained model for simulation of the DNA elasticity \cite{fathizadeh01,fathizadeh02}. The chain is considered as a system of successive rigid objects, where each object represents a base-pair. Each base-pair interacts with its nearest neighbors via a harmonic potential:
 
\begin{equation}
\label{Harmonic-Potential} U_{H}=\frac{1}{2}(\mathbf{\psi}-\mathbf{\psi}_0)^{T}\cdot\mathbf{K}\cdot(\mathbf{\psi}-\mathbf{\psi}_0)
\end{equation}

where $\mathbf{\psi}$ is a vector with 6 components, specifying relative orientation and separation of the adjacent base-pairs. The rotational (Twist, Tilt, and Roll) and translational (Shift, Slide and Rise) parameters are defined via CEHS representation \cite{El01,Olson01}. $\mathbf{\psi}_0$ represents the equilibrium configuration of base-pairs and $\mathbf{K}$ is a $6\times 6$ stiffness matrix.

A relatively good set of parameters~\cite{Becker01,Becker02}, which carries information from both all-atom molecular dynamics simulations~\cite{Lankas01} and protein-DNA crystal structures~\cite{Olson02}, have been employed here. Since in the current study DNA is long enough so that the sequence-dependent effects are averaged out, \textcolor{black}{we apply a homogeneous DNA and the effect of the DNA sequence is not considered}. So it is solely required to define one set of $\mathbf{\psi}_0$ and $\mathbf{K}$. We use the average values of the $\mathbf{K}$'s for all the ten possible sequence steps and we set all the components of the $\mathbf{\psi}_0$ to be equal to zero except for twist,  $Tw_0=35.0^\circ$, and rise, $Ri_0=0.34 nm$. 
\textcolor{black}{The above parameterization leads to a persistence length of about 150 base-pairs ($51$ nm) for the DNA.}
The above parameterization has the effect of short range electrostatic interactions implicitly. In the presence of monovalent ions, the long range interactions are screened in Debye length and they only have a short range contribution in repulsion, such that it affects the excluded volume interactions existing between the base-pairs. Then, in addition to the above harmonic potential, it is required to consider a self avoiding repulsive potential. This can be accomplished by considering ellipsoidal shape for each base-pair and implementing the repulsive part of the RE-squared potential. For two ellipsoids which interact via this potential we have \cite{Everaers01}
\begin{eqnarray}
\label{RE2-Potential} U^{RE-squared}_R=\frac{\mathbf{A}^{bb}}{2025}(\frac{\sigma ^{bb}_c}{h})^6(1+\frac{45}{56}\eta_{12}\chi_{12}\frac{\sigma ^{bb}_c}{h}) \times   \nonumber\\ 
\mathbf{\prod ^2_{i=1}} \ \mathbf{\prod _{e=x,y,z}}(\frac{\sigma _e^{(i)}}{\sigma _e^{(i)}+h/60^{1/3}}),
\end{eqnarray}
where $\mathbf{A}^{bb}=20 k_B T$ is the Hamaker constant for interaction between beads and $\sigma^{bb}_c=0.34 nm$ is the interaction radius. $\sigma _x^{(i)}$ and $\sigma _y^{(i)}$ and $\sigma _z^{(i)}$ are the half-radii of the ellipsoids. All of these parameters for interaction between the beads are set to the fitted values from an ellipsoidal rigid base-pair model of DNA \cite{Mergell01}. \textcolor{black}{These parameters lead to an effective diameter of about 2.0 nm for the DNA. But it is well known that the presence of water molecules hydrating the DNA and the screened electrostatic interaction of phosphates in presence of counterions can affect this value. To consider these effects we have set up simulations with $\sigma^{bb}_c=0.8 nm$ and $\sigma^{bb}_c=1.3 nm$ which make the effective diameter of the DNA to be about 2.5 nm and 3.0 nm respectively.} $h$ is the least distance between the two ellipsoids which is found by Gay-Berne approximation \cite{Gay01}. $\eta_{12}$ and $\chi_{12}$ depend on the orientation and separation of the ellipsoids and are calculated from two diagonal tensors, the structure tensor $\mathbf{S}_i=diag \lbrace \sigma _x,\sigma _y,\sigma _z \rbrace$, and the relative well-depth tensor $\mathbf{E}_i=\sigma_c \ diag \lbrace \frac{\sigma _x}{\sigma _y \sigma _z}, \frac{\sigma _y}{\sigma _x \sigma _z}, \frac{\sigma _z}{\sigma _x \sigma _y} \rbrace$ as
\begin{equation}
\label{ch12} 
\chi_{12}=2\hat{\mathbf{r}}_{12}^T \mathbf{B}_{12}^{-1} \hat{\mathbf{r}}_{12},
\end{equation}
and
\begin{equation}
\label{eta12} 
\eta_{12}=\frac{det[\mathbf{S}_1]/\sigma _1^2+det[\mathbf{S}_2]/\sigma _2^2}{[det[\mathbf{H}_{12}]/(\sigma _1 +\sigma _2)]^{1/2}},
\end{equation}
Here $\sigma_i=\hat{\mathbf{r}}_{12}^T \mathbf{R}_i^T \mathbf{S}_i^{-2} \mathbf{R}_i \hat{\mathbf{r}}_{12}$, where $\mathbf{R}_i$'s are the local orientational tensor of the $i^{th}$ ellipsoid and  $\hat{\mathbf{r}}_{12}$ is the relative position vector of the center of ellipsoids. The tensors $\mathbf{B}_{12}$ and $\mathbf{H}_{12}$ are defined as
\begin{equation}
\label{B12} 
\mathbf{B}_{12}=\mathbf{R}_{1}^T\mathbf{E}_{1}\mathbf{R}_{1}+\mathbf{R}_{2}^T\mathbf{E}_{2}\mathbf{R}_{2},
\end{equation}
and
\begin{equation}
\label{H12} 
\mathbf{H}_{12}=\frac{1}{\sigma _1}\mathbf{R}_1^T \mathbf{S}_{1}^2 \mathbf{R}_{1}+\frac{1}{\sigma _2}\mathbf{R}_2^T \mathbf{S}_{2}^2 \mathbf{R}_{2}.
\end{equation}

In the simulation this repulsive potential is only considered between those base-pairs which are separated from each other by more than 15 base-pairs along the chain. For the base-pairs which are closer to each other, the elastic energy cost, $U_H$, would play the role of self avoiding.

The spherical shell which confines the semiflexible chain is modeled by 1026 spherical particles with diameter of $1nm$, distributed uniformly on the surface of a sphere. The interaction between ellipsoids and spherical surface is also considered to be repulsive RE-squared potential with the same Hamaker constant, $\mathbf{A}^{bs}=\mathbf{A}^{bb}=20 k_B T$, and the interaction radius between them is supposed to be $3 \AA$. 

\subsection{Simulations}

Simulations are done through rigid body dynamics for the ellipsoidal beads and integrating the equations of motion using a symplectic algorithm \cite{Kamberaj01}. It is assumed that the mass of each base-pair is distributed uniformly in a geometry given in \cite{Mergell01} to find the corresponding moment of inertia. The repulsive forces and torques corresponding to RE-squared potential are calculated analytically \cite{Babadi01}. For the elastic part of the potential, $U_H$, forces are calculated analytically but the torques are calculated by the method of virtual work based on exerting small virtual rotations on each base-pair on three mutually perpendicular directions \cite{fathizadeh02}. The spherical nodes of the shell are fixed and have been excluded from the integration. The simulations are done in NVT ensemble at room temperature ($300$~K) and the temperature is controlled by Nose-Hoover chain thermostat during the simulations. 

\begin{figure}[h]

\includegraphics[width=7.0cm, angle=0]{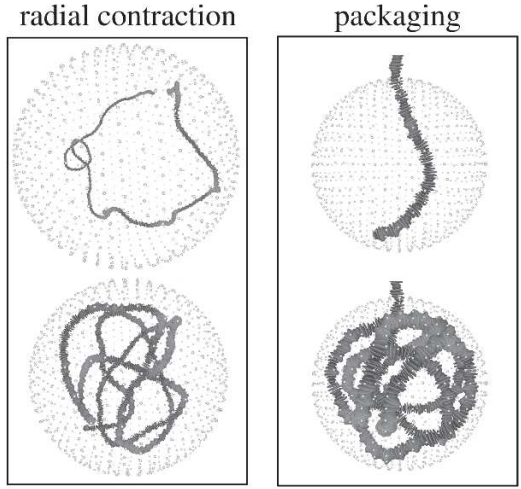}

\caption{Schematic representation of two simulation procedures to confine DNA chains. Panel A shows two snapshots of the first procedure which we put a semiflexible chain inside a sphere and then sphere gradually shrinks as long as it contains the semiflexible polymer. In the other procedure which is shown in panel B we try to pack the chain inside the sphere by injecting through a hole on the sphere.} \label{fig1}

\end{figure}

We studied conformations of a confined semiflexible chain with two different procedures. The first one is a radial contraction of the shell as shown schematically in the left panel of figure \ref{fig1}. At the beginning we generate a thermalized free chain of base-pairs by Monte Carlo method. This is achieved by picking the values of the base-pair step parameters from a Gaussian distribution according to the Boltzmann statistics. This sampling method helps us to obtain the equilibrium state of a long chain with a very short computational cost. After this step, the relaxed chain is located inside a spherical shell. The initial diameter of the shell is considered to be 1.1 times larger than the largest distance between the base-pairs along the relaxed chain. Starting the contraction of the shell, the radius of the shell is reduced regularly at each MD time step by a constant rate of about $\frac{1}{5000} \AA$ per timestep. The contraction is continued until the shell reaches its desired radius. After that the contraction is stopped and the system is equilibrated with performing MD while the shell radius is fixed. Then the conformation of the chain is sampled for time averaging and further investigations. The procedure is done for two given values of final radii, $8$ and $10$~nm. The final diameter of the spheres are approximately $2$ to $3$ times smaller than DNA persistence length which is roughly $48$~nm in our model. This will lead to a significant elastic energy cost for the polymer chain.

In the other approach we package the chain by injecting it into a spherical shell with a fixed radius. A hole is set on the shell surface by removing a few particles from the shell (5 and 12 particles for $8$ and $10$~nm spheres respectively). Again we begin with a chain of relaxed base-pairs, generated by Monte Carlo sampling. This time we insert the head of the polymer chain into the hole in a way that the first 3 base-pairs are located in the capsid initially. The chain is equilibrated at room temperature while its head is kept fixed. After this step a radial force of 50 pN toward the centers of the sphere is exerted on the center of the mass of all the base-pairs which are located in a cylindrical region of radius $1$~nm and height of $0.3$~nm above the hole. The force injects the chain inside the shell. The magnitude of the force is chosen to be close to the reported forces in the packaging process of DNA inside $\phi29$ and $\lambda$ bacteriophages \cite{Smith02,Fuller01,Fuller02}. As soon as the packaging process is completed, the system is equilibrated and then sampled for conformational analyses. We have done the packaging simulation for two given radii of the sphere as well. 
For both confinement scenarios, we repeated the simulations for 20 different realizations (10 for each sphere size) for a chain of 1000 base-pairs. Some movies from both confinement methods are available via this link http://softmatter.cscm.ir/polymer-confinement/index.htm . \textcolor{black}{For comparison we also tried simulations with 2000 base-pairs inside the sphere of $10$ nm.}

\section{Results and Discussion}
\begin{figure}[h]
 \includegraphics[width=9.0cm, angle=0]{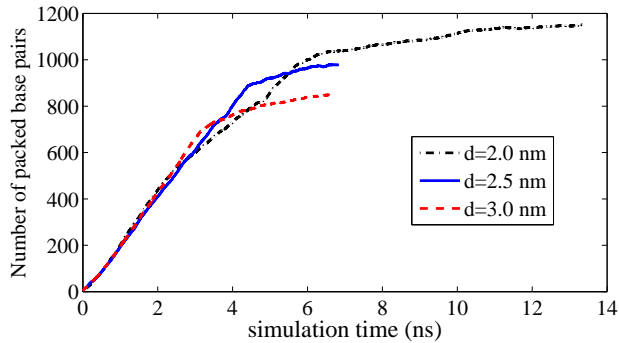}
\caption{Number of packed base-pairs versus time for a chain with 1500 base-pairs inside a sphere with $R=8nm$ for different polymer thicknesses. } 
\label{packaging-number}
\end{figure}

In order to estimate \textcolor{black}{the capacity of the capsids for different polymer thicknesses, we have tried to inject chains having a length of 1500 and 2500 base-pairs inside capsids of $8$ nm and $10$ nm respectively to find the ultimate value of packaging in these sizes. Figure }\ref{packaging-number} \textcolor{black}{shows the number of packed base-pairs versus time for DNA packaging inside $8$ nm sphere with $50$ pN insertion force found from three separate simulations of chains with three different diameters ($d=2.0,2,5,3.0$ nm). The simulations stop at the end of the packaging process when the capsid becomes full. According to the fact that the system is simulated in NVT ensemble, the thermostat damps the inertial forces and from this figure one can see that initially the packaging starts with a constant rate regardless of the diameter of the chain. This rate slows down when the capsid becomes partially full and the energetic forces resist against the packaging process. These simulations are repeated 3 times for every capsid and every polymer thickness and the capacity of the capsid is reported by averaging over the 3 performed simulations. The results are shown in table} \ref{capacity}. \textcolor{black}{It is obvious that for thicker polymers the capacity reduces. Also one can see that the capacity of the $10$ nm sphere is approximately two times bigger than the $8$ nm sphere. }
 \textcolor{black}{In our simulations we observed that the diameter of the polymer in these sizes does not have a significant effect on the conformation of the confined polymer so from now on we only discuss about the conformations of the polymer with diameter of $2.0$ nm.} The ultimate amount of the packed chain with this diameter is about 1070-1170 base-pairs for this setup. So choosing a chain with 1000 base-pairs allows us to obtain about $85\%$ of the ultimate capacity at $8$ nm sphere size. Another fact which can be seen from one simulation of packaging of the polymer with diameter of $2.0$ nm  (figure \ref{packaging-number}) is that for the first 1000 base-pairs of the packed chain, the packaging rate is almost constant (the graph is linear) and after that, there is a crossover and the packaging process slows down. Indeed, the next 150 base-pairs are packed in almost the same simulation time as for the first 1000 base-pairs. So by choosing a chain with 1000 base-pairs in length, the simulations do still remain fast enough to reduce the computational costs while a nearly full ($85\%$) capsid is achieved. From the table \ref{capacity}, one can observe that just about $40\%$ percent of the volume is filled by the 1000 base-pairs polymer chain.

\begin{table}[h]
\caption{The capacity of spheres for 3 values of polymer diameter (d) }
\centering                         
\begin{tabular}{|c|c|c|}     
\hline 
 \ \ $d (nm)$ \ \ & \ \  $D=16 nm$ \ \ & \ \ $D=20 nm$ \ \  \\
\hline 
 \ \ 2.0 \ \   & \ \ 1120 $\pm$ 55 \ \ &  \ \ 2378 $\pm$ 71 \ \ \\
\hline
 \ \ 2.5 \ \ &  \ \ 935 $\pm$ 36 \ \ &  \ \ 1882 $\pm$ 47 \ \ \\
\hline                           
 \ \ 3.0 \ \  &  \ \ 848 $\pm$ 28 \ \ &  \ \ 1794 $\pm$ 33 \ \ \\
\hline
\end{tabular}
\label{capacity}          
\end{table}

\begin{figure}[h]

 \includegraphics[width=7.5cm, angle=0]{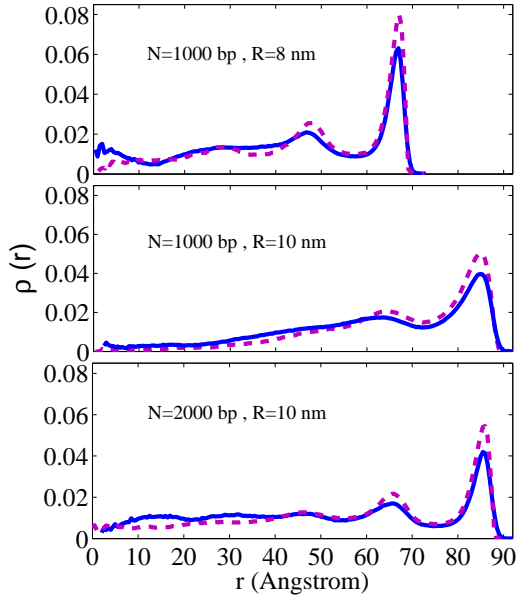}
 
\caption{(Color online) Radial distribution of density of the base-pairs of the packed chain for packaging (blue solid curves) and for radial contraction (dashed cyan curves) for sphere sizes of $R= 8 $nm and $R= 10 $nm and for chains with 1000 and 2000 base-pairs as it is shown on the figure. }
\label{density}
\end{figure}

\textcolor{black}{To analyze the structure of the confined polymer, first we need to investigate how base-pairs are distributed inside the sphere in either confinement methods. }Figure \ref{density} \textcolor{black}{shows the distribution of the density of 1000 base-pairs, $\rho (r)$, as a function of distance from the center of spheres for two methods and for two given sphere sizes. For the bigger sphere, the distribution is plotted also for 2000 base-pairs which has the same density as 1000 base-pairs inside the $8$ nm sphere. }

The values of $\rho (r)$ are averaged over several realizations and also over time for each simulation. It also normalized in a way that $\int \rho(r) dr=1$.  For both methods and both radii, some peaks appear in the plot of $\rho (r)$. \textcolor{black}{Such layered structures have been observed in previous studies of DNA packaging} \cite{Spakowitz01,Rollins01}. The peaks are about 2nm apart which is approximately the size of the double stranded DNA thickness. The highest peaks appear near the surface of the sphere and the reason is that the semiflexible chain tends to maximize its radius of curvature to reduce the bending elastic energy. For $8$~nm sphere, the onion-like layered structure is more visible in comparison with $10$~nm sphere where the structure is fade out after second layer. The small appeared peak near the center of the sphere in the packaging method is due the fact that in this procedure the base-pairs are fed into the sphere toward its center. \textcolor{black}{In the simulation performed by Spakowitz and Wang} \cite{Spakowitz01}, \textcolor{black}{the appearance of peaks near the sphere surface as well as the center of the sphere has been reported and it's mentioned that the former is the highest one. In contrast, another study by Rollins et al. \cite{Rollins01} proposes that the peak near the center of the sphere is  the highest one.} In radial contraction method, near surface peak is more significant. This clearly shows the different dynamics in the two confining methods; in the radial contraction, the polymer chain is confined by an inward radial force, while in the packaging method, it is fed to a sphere from the center. \textcolor{black}{The behavior of the curves for both methods with 2000 base-pairs are very close to those obtained for 1000 base-pairs inside the sphere with $R= 8 $nm.}

\begin{figure}[h]

\includegraphics[width=6.0cm, angle=0]{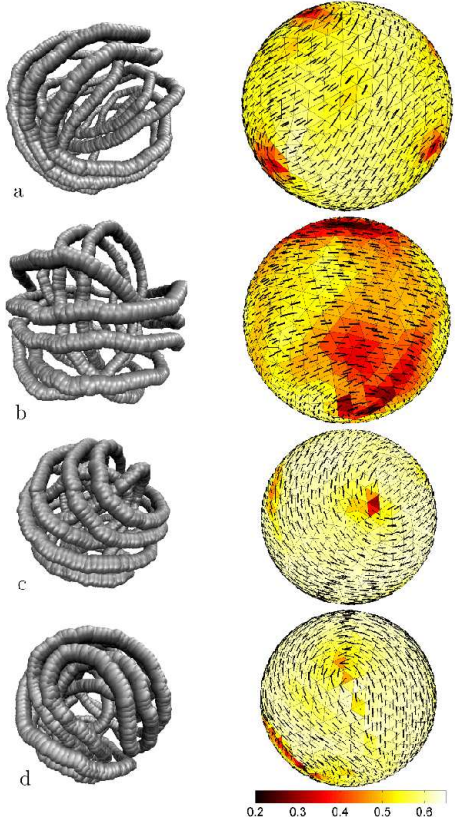}

\caption{(Color online) Some snapshots of the structure of the confined chain inside the spheres (left) and their corresponding nematic order analysis (right) obtained in the radial contraction procedure. Panels (a) and (b) show relaxed configurations for the confined chain in spheres with $r_0=10nm$ in two separate simulations. (a) shows a tennis-ball configurations and configuration in (b) is spool shape. Panels (c) and (d) show DNAs in (a) and (b), when the contraction is continued to reach a $r_0=8nm$ sphere.} \label{gradual}

\end{figure}

Figure~\ref{gradual} presents some snapshots of the simulations on DNA with 1000 base-pairs to demonstrate structures visually. In the left column of the figure, structures of the chain obtained in radial contraction are shown for two given radii. Panels (a) and (b) correspond to the conformations of the chain confined in a sphere with radius of $10nm$. As one can realize in panel (a) the chain has a tennis-ball structure and this is very likely such that 8 out of 10 simulations result this conformation. The other observed conformation which is illustrated in panel (b) has spool-like structures. So the tennis-ball structure of the chain is more probable under radial contraction dynamics. This observation is more justified if the contraction is continued and the conformations of the chain are investigated in $8$~nm spheres. Panel (c) and (d) of figure \ref{gradual} show two examples of the obtained configurations. We found that all ten simulations lead to the tennis-ball conformation in the $8$ nm sphere. This is interesting because even the two observed spool-like structures in the $10$~nm spheres were disappeared by continuing the confinement to the $8$~nm. For instance, panel (d) in figure \ref{gradual} shows the evolution of the structure shown in panel (b), when the radius of the sphere is decreased from $10$ nm to $8$ nm. It can be seen that the spool structure is converted to the tennis-ball structure. This change in the conformation happens because of the buckling of the circular structures of the chain. The polymer circles are structured near the surface of the sphere and confined between the sphere surface and the second layer of the base-pairs as mentioned in figure \ref{density}. Consequently, continuing the confinement procedure causes the circular structures to buckle and form the folded toroids. A movie from this buckling mechanism is available via this link http://softmatter.cscm.ir/polymer-confinement/index.htm .

The structures of the confined chain can be clarified better if the concept of nematic order is employed in analogy with liquid crystals. Here the bond vector, the unit vector which connects centers of two successive base-pairs, is a proper choice for defining microscopic director. For those base-pairs which are close to the sphere surface, this vector tends to be tangent to the surface. This makes a good analogy with liquid crystal droplets, when the surface anchoring is homogeneous (tangential).  To create the director field on the surface of the sphere, we used nearest base-pairs to the spherical surface. This is done by choosing the base-pairs sitting in the closest layer of the density distribution to the surface (see figure~\ref{density}). Base-pairs which are in that shell are used for nematic order analysis. We take samples from our simulations in each sphere size and plot the average of normalized projection of the bond vectors for base-pairs near the surface. In this way, we achieve a director field which approximately covers the surface of the sphere. So the surface is meshed by small triangles and in each cell a local nematic order parameter tensor, $Q$, is defined as \cite{Kleman01},
\begin{equation}
\label{tensor} 
Q_{ij}=\frac{1}{N}\sum_{\alpha=1}^{N} (\frac{3}{2} u_{\alpha i}u_{\alpha j} -\frac{1}{2} \delta _{ij}),
\end{equation} 
where $u_{\alpha n}$ is the $n$th component of the $\alpha$th vector, $\delta$ is Kronecker delta function, and $N$ is the number of vectors contributing in the calculation of the nematic order parameter tensor in each triangular cell. The biggest eigenvalue of $Q$ tensor is proportional to the value of the scalar nematic order parameter and its corresponding eigenvector indicates the nematic director \cite{Kleman01}, in each element. In figure \ref{gradual}, on the right hand side of each panel, the maps of the nematic are shown. The vector field shows the local nematic director and the scalar order parameter is shown by color scheme. For tennis-ball structures four poles (defects) of $+1/2$ charge can be recognized. On the other hand, for the spool structures two $+1$ defects appear on the top and the bottom of the spool, where the nematic director revolves around (figure~\ref{gradual}b).  

\begin{figure}[h]

\includegraphics[width=6.0cm, angle=0]{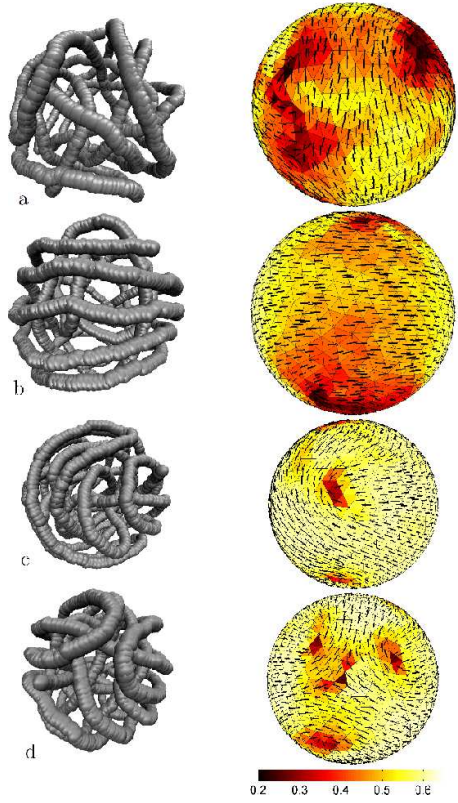}

\caption{(Color online) The structures of confined polymer chain inside the sphere (left column) and nematic director filed (right column) in packaging process. Each configuration corresponds to a separate simulation. Panels (a) and (b) correspond to packaging in spheres with radius of $10$~nm and panels (c) and (d) correspond to the packaging inside spheres with $8$~nm in radius.} \label{packaging}

\end{figure}

The same method is used to analyze the structure obtained in the packaging simulations. In figure \ref{packaging}, some of the resultant conformations of the packed chain as well as the corresponding nematic director fields are illustrated. Panel (a) shows conformation of the chain inside a sphere with $10$~nm radius. It can be seen that the chain is less ordered in comparison with the radial contraction process. This is also visible in the director map, which shows a combination of both tennis-ball and spool-like conformations. This happened in 8 out of 10 packaging simulations for this sphere size. In the 2 remaining simulations, the chain tends to form a spool structures. A sample snapshot is shown in panel (b) of figure \ref{packaging}. Panel (c) and (d) in this figure show two snapshots of two packaging simulations for the $8$~nm sphere. In this size, the tennis-ball conformations (panel (c)) are more likely and have been observed $7$ times in different realizations. Although the tennis-ball structure is recognizable in both figures of panel (c), the structures are not that perfect as they are in the radial contraction (panel (c) and (d) of figure \ref{gradual}). On the other hand, in 3 out of 10 simulations we found less ordered structures with higher defect excitations. For example the one is shown in panel (d), has an irregular shape with a $-1/2$ defect on the surface, however, the total charge is $+2$ and obeys Poincare-Hopf theorem.

\begin{figure}[h]

\includegraphics[width=5.0cm, angle=0]{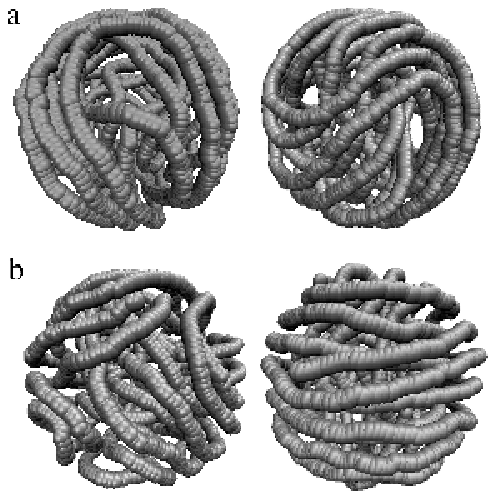}

\caption{(Color online) \textcolor{black}{Some observed structures of the confined chain with 2000 base-pairs inside the sphere with $R=10 nm$ for radial contraction (panel a) and packaging (panel b). Each of the four configurations corresponds to a separate simulation. Radial contraction always leads to tennis-ball structure for this density. The disordered conformations are the most likely structures for packaging at this size but spool-like structures sometimes can be obsereved.}}  \label{2000all}
\end{figure}

Figure \ref{2000all} \textcolor{black}{shows some snapshots of the simulation of a chain with 2000 base-pairs in sphere with $R=10 nm$ for radial contraction (top) and packaging (bottom). Similarly, we still mostly observe tennis-ball structures in radial contraction and for packaging the disordered structures are the most likely while spool-like structures sometimes occur.}

It is interesting to compare the elastic energy costs for the polymer chains in two methods. The elastic energy is obtained from equation~(\ref{Harmonic-Potential}). Every base-pair (bead) has three rotational and three translational degrees of freedom. So a thermalized free chain with 1000bp's gets $U_{eq}=1000\times{6\over2} k_B T$ amount of elastic energy because of thermal fluctuations. The difference between elastic energy of the confined chain and a free one, $\Delta U_{el}$, will give us the elastic energy cost for the confinement. This calculation also can be done separately for different parts of the elastic energy such as bending, twisting, and stretching. In table \ref{energies} the total elastic energy and the bending energy, $\Delta U_{bend}$, for both confinement methods and both sphere sizes are shown. 
     
The table shows that in the both confinement methods almost the whole amount of the confinement elastic energy comes from bending energy and the other degrees of freedom are almost relaxed. There is no significant energy difference in other components of the energy (not shown in the table). Comparing two methods, it can be seen that energies of the polymer chain in radial contraction are less than those in the packaging simulations and this difference becomes larger for smaller sphere (about $60~k_{\rm B}T$ for the sphere with radius of $8$~nm). Thus on average, the radial contraction will lead to the more energetically relaxed structures. In the last column of the table~\ref{energies}, the average amount of the change in twist in the confined chain with respect to a free chain is mentioned. This value can be easily obtained by a summation over twist values of all base-pair steps (using CEHS definition \cite{El01,Olson01}) and averaging over time. These results show that the excess amount of twist is not significant in either method.

\begin{table}[h]
\caption{The confinement elastic energy ($k_{\rm B}T$) and the change in twist of the polymer chain.}
\centering                         
\begin{tabular}{|c|c|c|c|c|}     
\hline  \hline      
\ \ & $r (nm)$ & \  $\Delta U_{el}$ \ & \ $\Delta U_{bend}$  \ &  \ $\Delta Tw$  \ \\
\hline 
 radial  & $10$  & 175.3 $\pm$ 4.9 &  181.0 $\pm$ 4.4 &  -0.20 $\pm$ 0.04\\
 contraction & $8$   &  363.1 $\pm$ 5.3 &  367.0 $\pm$ 5.0 & -0.38 $\pm$ 0.08\\
\hline
 packaging & $10$  &  199.0 $\pm$ 10.5 &  190.1 $\pm$ 9.2 & -0.18 $\pm$ 0.05 \\
 & $8$  &  422.7 $\pm$ 27.7  &  417.6 $\pm$ 24.3 &  -0.68 $\pm$ 0.15 \\
\hline                           
\end{tabular}
\label{energies}          
\end{table}

\begin{figure}[h]

\includegraphics[width=7.0cm, angle=0]{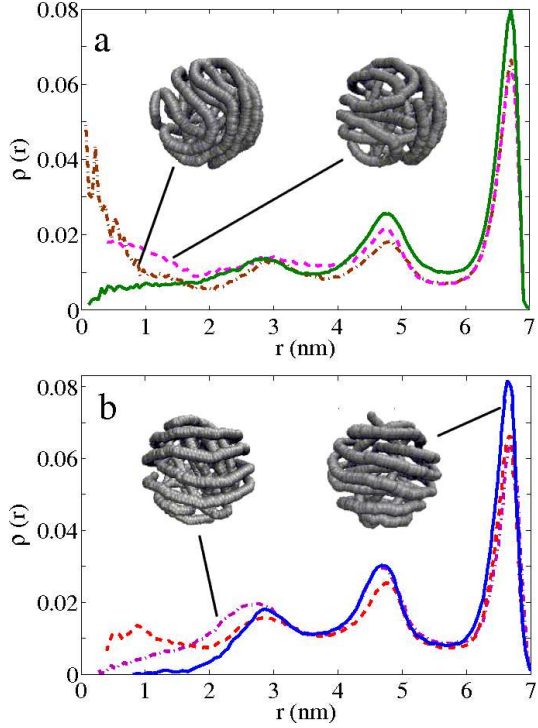}

\caption{(Color online) Comparison of density distribution for different confinement rates.  (a) compares density distribution of the confined polymer in a radial contraction with rates $2$ times (dot-dashed brown) and $4$ times (dashed pink) faster than the reported rate of $\frac{1}{5000} \AA$ (solid green). (b) compares packaging by 100 pN (dot-dashed pink) and 150 pN (solid blue) insertion forces in $8nm$ sphere with $50$~pN (dashed red). The corresponding conformations of fast processes are shown schematically as insets.}
\label{fast-confinement}
\end{figure}

To be sure that the results are not subjected by the confinement rate, the above presented simulations are also performed with slower procedures. In radial contraction, the confining rate has been slowed down by a factor of 4 and for insertion packaging, the packaging forces of $35$, $40$, and $45$~pN have been tried. In either method no significant differences in results have been observed for slower dynamics. On the other hand, when we make the process faster some changes in configurations are observed, while the distribution of the base-pairs inside the spheres remains almost the same as before with onion-shell structure. Figure~\ref{fast-confinement} compares distribution of base-pairs inside $8$~nm sphere for three different rates of confinement. For the radial contraction (figure~\ref{fast-confinement}a) solid curve shows the distribution of the base-pairs with rate of $\frac{1}{5000} \AA$ and the contraction rates of dot-dashed and dotted curves are two and four times faster. For packaging simulation we double and triple packaging force (dot-dashed and solid lines in figure~\ref{fast-confinement}b). As it can be seen, in either case the onion-like structures of density distribution are preserved and they only differ in the center. Regarding the structures, in the radial contraction the tennis-ball conformation is still observed, although the structure becomes less regular at high contraction rates (see the figure \ref{fast-confinement}a insets) . On the contrary, in the packaging with high packaging forces, the spool-like configuration is the dominant configuration. Formation of the circles of the polymer in packaging at small sizes requires spending much more energy. This energy cannot be achieved unless the insertion force becomes large enough. So in small capsids with small packaging forces the tennis ball conformation is the preferred structure, but by increasing the forces the conformation transforms to the spool-like structure (see the figure\ref{fast-confinement}b insets). Two movies from packaging with 50 pN and 100 pN injection forces are available via this link http://softmatter.cscm.ir/polymer-confinement/index.htm . 

\section{Effect of electrostatic attraction}
\begin{figure}[h]

\includegraphics[width=5.0cm, angle=0]{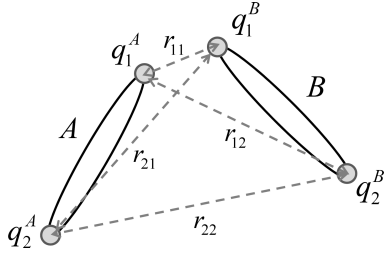}

\caption{The electrostatic interaction between two base-pairs. Ellipses A and B represent the base-pairs. The electrostatic attraction consists of four interactions between two end points of the base-pairs.}
\label{electrostatic}
\end{figure}

For a DNA chain, the electrostatic interaction due to the negatively charged phosphate groups along the DNA chain affects its properties in several ways. One part of this effect on the DNA elastic parameters is already included in the elastic potential parameters of neighboring base-pairs (equation \ref{Harmonic-Potential})employed in this study. But in the presence of the multivalent counterions, this interaction may transform into an attraction between two DNA double strands \cite{Bloomfield01}.In the simulations of DNA packaging, this effect is usually considered as a short range interaction between the base-pairs \cite{Kindt01,Ali01, Marenduzzo01, Sottas01}.  The elastic part of the energy usually dominates over the electrostatic part in determining the structure of DNA inside the capsid \cite{angelescu03,Petrov02}. However, to show that our results can be applicable to the DNA packaging problem, here we will examine the effect of electrostatic attraction in our model by introducing a short range attractive potential. In the rigid base-pair chain model that the phosphate groups are absent, this short range attraction between the phosphates can be approximated with a Van der Waals potential between the end points of the base-pairs (See figure \ref{electrostatic}.):

\begin{equation}
\label{LJ} 
U_{VDW} = \left\{ \begin{array}{rl}
 \sum_{i=1}^{2} \sum_{j=1}^{2} 4 \epsilon ((\frac{\sigma}{r_{ij}})^{12}-(\frac{\sigma}{r_{ij}})^{6}) &\mbox{ if $r_{ij}\leq 2.5\sigma$} \\
  0 \ \ \ \ \ \ \ \ \ \ \ \ \ \ \ \ \ \ \ \ \ \ \ \ \ \ \ \ \ \ \ \ \ \ \ \ \ \ \ \ \ &\mbox{ otherwise}
       \end{array} \right.
\end{equation}

\begin{figure}[h]

\includegraphics[width=6.0cm, angle=0]{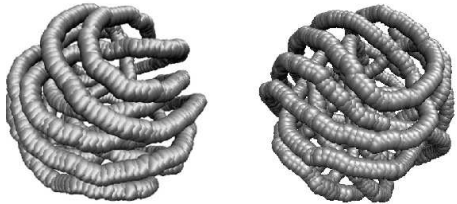}

\caption{Samples of the chain conformations in the packaging (right) and radial contraction (left) in the presence of attractive interaction with $\epsilon=1.0 k_T$ in a sphere with $8nm$ radius.}
\label{attractive}
\end{figure}

In the above equation, $r_{ij}$'s are the distances between the end points and $\sigma=0.25 nm$. For the depth of the potential, $\epsilon$, usually a range between 0.4-0.8 $k_BT$ have been used in the literature \cite{Kindt01,Ali01, Marenduzzo01, Sottas01}. The above potential acts on every two base-pairs that at least have one $r_{ij}$ inside the cutoff ($2.5\sigma$). To see whether this interaction may affect the dynamics of the confinement in our study, we set up simulations with the attraction energy according to equation \ref{LJ} with two $\epsilon$ values of $0.6 k_BT$ and $1.0 k_BT$. A sample of the obtained conformations is shown in figure \ref{attractive} for sphere size of $8 nm$ and $\epsilon=1.0 k_BT$. As it is shown, this attraction is not strong enough to have an influence on the dynamics of confinement and change the chain conformation. Still the tennis-ball structure is the favored structure in the radial contraction and also a signature of this structure can be observed in the packaging at this sphere size. Calculations of the energies show that the total attractive energies of the confined DNA chains for $\epsilon=0.6 k_BT$ and $\epsilon=1.0 k_BT$ are about $-55 k_BT$ and $-100 k_BT$ respectively. The attraction energy acts against the bending energy, holds the DNA close together and consequently, reduces the pressure on the sphere's surface. Although the attractive energy does not have a significant effect on the DNA conformation and dynamics in our study, this effect can play a role on the stability of the structure. 

\section{Conclusions}
The current study of the confined semiflexible polymer chains in a sphere, proposed some new interesting aspects of conformation of the polymers in extreme confinements. Using a coarse grained molecular dynamics simulation, we studied two different dynamical procedures for confining the polymer chain inside spheres of two different sizes. Although the scales of the confinement in our studies are smaller than usual viral capsid sizes, the results can be helpful to understand DNA structure inside viral capsids. \textcolor{black}{Our packaging results can be useful to understand the conformation of the DNA at the final stages of the packaging in viral capsids where the elastic energy becomes very important and has a significant role on conformation of the internal part of the packed DNA.} In summary, our study proposes that the conformation of a semiflexible chain in spherical confinement significantly depend on the dynamics of the confinement and confining procedure. Different dynamics or different rates of confinement leaded to different conformation for the chain. We mostly focused on the conformation of DNA near the surface of the sphere. 

In the radial contraction procedure, the tennis ball structures are the dominant observed conformation for the confined chain in nanometer sized capsids. Even though the spool conformation sometimes observed in the sphere of $10$~nm radius, by resuming the contraction procedure these structures tend to the tennis ball because of buckling of the polymer chain circles. 

In the packaging procedure for $10$~nm sphere size mostly leaded to the unarranged or mixed conformations but in a few simulations a weak signature of the spool structure observed. However, the tennis ball structure has never been reported in this sphere scale. Like the radial contraction, in $8nm$ sphere the tennis ball structure were the most observed structure, although, higher excitations have been observed. \textcolor{black}{Our results show that the proposed tennis-ball structure for confined polymer in spherical confinement by Katzav et al. } \cite{Katzav01}, \textcolor{black}{can be observed in confined structures of DNA inside small nano-spheres in which the elastic energy plays an important role. The inner layers of the packed DNA  in viral capsids are good candidates to look for such tennis-ball structures.} 

Iit was shown that applying a short ranged interaction between the base-pairs due to the electrostatic effect of counterions on the phosphates of the DNA chain does not affect its structure in confinement while can be important on the stability of the structure and capsid. Also increasing the packaging force changes the packaging mechanism and pushs the chain to form the spool structures.

\section*{Acknowledgement}
We wish to deeply thank Prof. Mehdi Habibi for helpful discussions.


\begin{thebibliography}{0}%
\makeatletter
\providecommand \@ifxundefined [1]{%
 \@ifx{#1\undefined}
}%
\providecommand \@ifnum [1]{%
 \ifnum #1\expandafter \@firstoftwo
 \else \expandafter \@secondoftwo
 \fi
}%
\providecommand \@ifx [1]{%
 \ifx #1\expandafter \@firstoftwo
 \else \expandafter \@secondoftwo
 \fi
}%
\providecommand \natexlab [1]{#1}%
\providecommand \enquote  [1]{``#1''}%
\providecommand \bibnamefont  [1]{#1}%
\providecommand \bibfnamefont [1]{#1}%
\providecommand \citenamefont [1]{#1}%
\providecommand \href@noop [0]{\@secondoftwo}%
\providecommand \href [0]{\begingroup \@sanitize@url \@href}%
\providecommand \@href[1]{\@@startlink{#1}\@@href}%
\providecommand \@@href[1]{\endgroup#1\@@endlink}%
\providecommand \@sanitize@url [0]{\catcode `\\12\catcode `\$12\catcode
  `\&12\catcode `\#12\catcode `\^12\catcode `\_12\catcode `\%12\relax}%
\providecommand \@@startlink[1]{}%
\providecommand \@@endlink[0]{}%
\providecommand \url  [0]{\begingroup\@sanitize@url \@url }%
\providecommand \@url [1]{\endgroup\@href {#1}{\urlprefix }}%
\providecommand \urlprefix  [0]{URL }%
\providecommand \Eprint [0]{\href }%
\providecommand \doibase [0]{http://dx.doi.org/}%
\providecommand \selectlanguage [0]{\@gobble}%
\providecommand \bibinfo  [0]{\@secondoftwo}%
\providecommand \bibfield  [0]{\@secondoftwo}%
\providecommand \translation [1]{[#1]}%
\providecommand \BibitemOpen [0]{}%
\providecommand \bibitemStop [0]{}%
\providecommand \bibitemNoStop [0]{.\EOS\space}%
\providecommand \EOS [0]{\spacefactor3000\relax}%
\providecommand \BibitemShut  [1]{\csname bibitem#1\endcsname}%
\let\auto@bib@innerbib\@empty
\end{thebibliography}%


\begin{thebibliography}{44}%
\makeatletter
\providecommand \@ifxundefined [1]{%
 \@ifx{#1\undefined}
}%
\providecommand \@ifnum [1]{%
 \ifnum #1\expandafter \@firstoftwo
 \else \expandafter \@secondoftwo
 \fi
}%
\providecommand \@ifx [1]{%
 \ifx #1\expandafter \@firstoftwo
 \else \expandafter \@secondoftwo
 \fi
}%
\providecommand \natexlab [1]{#1}%
\providecommand \enquote  [1]{``#1''}%
\providecommand \bibnamefont  [1]{#1}%
\providecommand \bibfnamefont [1]{#1}%
\providecommand \citenamefont [1]{#1}%
\providecommand \href@noop [0]{\@secondoftwo}%
\providecommand \href [0]{\begingroup \@sanitize@url \@href}%
\providecommand \@href[1]{\@@startlink{#1}\@@href}%
\providecommand \@@href[1]{\endgroup#1\@@endlink}%
\providecommand \@sanitize@url [0]{\catcode `\\12\catcode `\$12\catcode
  `\&12\catcode `\#12\catcode `\^12\catcode `\_12\catcode `\%12\relax}%
\providecommand \@@startlink[1]{}%
\providecommand \@@endlink[0]{}%
\providecommand \url  [0]{\begingroup\@sanitize@url \@url }%
\providecommand \@url [1]{\endgroup\@href {#1}{\urlprefix }}%
\providecommand \urlprefix  [0]{URL }%
\providecommand \Eprint [0]{\href }%
\providecommand \doibase [0]{http://dx.doi.org/}%
\providecommand \selectlanguage [0]{\@gobble}%
\providecommand \bibinfo  [0]{\@secondoftwo}%
\providecommand \bibfield  [0]{\@secondoftwo}%
\providecommand \translation [1]{[#1]}%
\providecommand \BibitemOpen [0]{}%
\providecommand \bibitemStop [0]{}%
\providecommand \bibitemNoStop [0]{.\EOS\space}%
\providecommand \EOS [0]{\spacefactor3000\relax}%
\providecommand \BibitemShut  [1]{\csname bibitem#1\endcsname}%
\let\auto@bib@innerbib\@empty
\bibitem [{\citenamefont {Marenduzzo}\ \emph {et~al.}(2010)\citenamefont
  {Marenduzzo}, \citenamefont {Micheletti},\ and\ \citenamefont
  {Orlandini}}]{Marenduzzo03}%
  \BibitemOpen
  \bibfield  {author} {\bibinfo {author} {\bibfnamefont {D.}~\bibnamefont
  {Marenduzzo}}, \bibinfo {author} {\bibfnamefont {C.}~\bibnamefont
  {Micheletti}}, \ and\ \bibinfo {author} {\bibfnamefont {E.}~\bibnamefont
  {Orlandini}},\ }\href@noop {} {\bibfield  {journal} {\bibinfo  {journal} {J.
  Phys.: Cond. Matt.}\ }\textbf {\bibinfo {volume} {22}},\ \bibinfo {pages}
  {283102} (\bibinfo {year} {2010})}\BibitemShut {NoStop}%
\bibitem [{\citenamefont {Elrad}\ and\ \citenamefont {Hagan}(2010)}]{Elrad01}%
  \BibitemOpen
  \bibfield  {author} {\bibinfo {author} {\bibfnamefont {O.}~\bibnamefont
  {Elrad}}\ and\ \bibinfo {author} {\bibfnamefont {M.}~\bibnamefont {Hagan}},\
  }\href@noop {} {\bibfield  {journal} {\bibinfo  {journal} {Phys. Biol.}\
  }\textbf {\bibinfo {volume} {7}},\ \bibinfo {pages} {045003} (\bibinfo {year}
  {2010})}\BibitemShut {NoStop}%
\bibitem [{\citenamefont {Morrison}\ and\ \citenamefont
  {Thirumalai}(2009)}]{Morrison01}%
  \BibitemOpen
  \bibfield  {author} {\bibinfo {author} {\bibfnamefont {G.}~\bibnamefont
  {Morrison}}\ and\ \bibinfo {author} {\bibfnamefont {D.}~\bibnamefont
  {Thirumalai}},\ }\href@noop {} {\bibfield  {journal} {\bibinfo  {journal}
  {Phys. Rev. E}\ }\textbf {\bibinfo {volume} {79}},\ \bibinfo {pages} {011924}
  (\bibinfo {year} {2009})}\BibitemShut {NoStop}%
\bibitem [{\citenamefont {Liu}\ and\ \citenamefont
  {Chakraborty}(2008)}]{Liu01}%
  \BibitemOpen
  \bibfield  {author} {\bibinfo {author} {\bibfnamefont {Y.}~\bibnamefont
  {Liu}}\ and\ \bibinfo {author} {\bibfnamefont {B.}~\bibnamefont
  {Chakraborty}},\ }\href@noop {} {\bibfield  {journal} {\bibinfo  {journal}
  {Phys. Biol.}\ }\textbf {\bibinfo {volume} {5}},\ \bibinfo {pages} {026004}
  (\bibinfo {year} {2008})}\BibitemShut {NoStop}%
\bibitem [{\citenamefont {Cerd{\`a}}\ \emph {et~al.}(2005)\citenamefont
  {Cerd{\`a}}, \citenamefont {Sintes},\ and\ \citenamefont
  {Chakrabarti}}]{Cerda01}%
  \BibitemOpen
  \bibfield  {author} {\bibinfo {author} {\bibfnamefont {J.}~\bibnamefont
  {Cerd{\`a}}}, \bibinfo {author} {\bibfnamefont {T.}~\bibnamefont {Sintes}}, \
  and\ \bibinfo {author} {\bibfnamefont {A.}~\bibnamefont {Chakrabarti}},\
  }\href@noop {} {\bibfield  {journal} {\bibinfo  {journal} {Macromolecules}\
  }\textbf {\bibinfo {volume} {38}},\ \bibinfo {pages} {1469} (\bibinfo {year}
  {2005})}\BibitemShut {NoStop}%
\bibitem [{\citenamefont {Ali}\ \emph {et~al.}(2004)\citenamefont {Ali},
  \citenamefont {Marenduzzo},\ and\ \citenamefont {Yeomans}}]{Ali02}%
  \BibitemOpen
  \bibfield  {author} {\bibinfo {author} {\bibfnamefont {I.}~\bibnamefont
  {Ali}}, \bibinfo {author} {\bibfnamefont {D.}~\bibnamefont {Marenduzzo}}, \
  and\ \bibinfo {author} {\bibfnamefont {J.}~\bibnamefont {Yeomans}},\
  }\href@noop {} {\bibfield  {journal} {\bibinfo  {journal} {J. Chem. Phys.}\
  }\textbf {\bibinfo {volume} {121}},\ \bibinfo {pages} {8635} (\bibinfo {year}
  {2004})}\BibitemShut {NoStop}%
\bibitem [{\citenamefont {Micheletti}\ \emph {et~al.}(2011)\citenamefont
  {Micheletti}, \citenamefont {Marenduzzo},\ and\ \citenamefont
  {Orlandini}}]{Micheletti01}%
  \BibitemOpen
  \bibfield  {author} {\bibinfo {author} {\bibfnamefont {C.}~\bibnamefont
  {Micheletti}}, \bibinfo {author} {\bibfnamefont {D.}~\bibnamefont
  {Marenduzzo}}, \ and\ \bibinfo {author} {\bibfnamefont {E.}~\bibnamefont
  {Orlandini}},\ }\href@noop {} {\bibfield  {journal} {\bibinfo  {journal}
  {Phys. Rep.}\ }\textbf {\bibinfo {volume} {504}},\ \bibinfo {pages} {1}
  (\bibinfo {year} {2011})}\BibitemShut {NoStop}%
\bibitem [{\citenamefont {Petrov}\ and\ \citenamefont
  {Harvey}(2008)}]{Petrov02}%
  \BibitemOpen
  \bibfield  {author} {\bibinfo {author} {\bibfnamefont {A.}~\bibnamefont
  {Petrov}}\ and\ \bibinfo {author} {\bibfnamefont {S.}~\bibnamefont
  {Harvey}},\ }\href@noop {} {\bibfield  {journal} {\bibinfo  {journal}
  {Biophys. J.}\ }\textbf {\bibinfo {volume} {95}},\ \bibinfo {pages} {497}
  (\bibinfo {year} {2008})}\BibitemShut {NoStop}%
\bibitem [{\citenamefont {Ali}\ \emph {et~al.}(2008)\citenamefont {Ali},
  \citenamefont {Marenduzzo},\ and\ \citenamefont {Yeomans}}]{Ali01}%
  \BibitemOpen
  \bibfield  {author} {\bibinfo {author} {\bibfnamefont {I.}~\bibnamefont
  {Ali}}, \bibinfo {author} {\bibfnamefont {D.}~\bibnamefont {Marenduzzo}}, \
  and\ \bibinfo {author} {\bibfnamefont {J.}~\bibnamefont {Yeomans}},\
  }\href@noop {} {\bibfield  {journal} {\bibinfo  {journal} {Biophys. J.}\
  }\textbf {\bibinfo {volume} {94}},\ \bibinfo {pages} {4159} (\bibinfo {year}
  {2008})}\BibitemShut {NoStop}%
\bibitem [{\citenamefont {Marenduzzo}\ \emph {et~al.}(2009)\citenamefont
  {Marenduzzo}, \citenamefont {Orlandini}, \citenamefont {Stasiak},
  \citenamefont {Sumners}, \citenamefont {Tubiana},\ and\ \citenamefont
  {Micheletti}}]{Marenduzzo01}%
  \BibitemOpen
  \bibfield  {author} {\bibinfo {author} {\bibfnamefont {D.}~\bibnamefont
  {Marenduzzo}}, \bibinfo {author} {\bibfnamefont {E.}~\bibnamefont
  {Orlandini}}, \bibinfo {author} {\bibfnamefont {A.}~\bibnamefont {Stasiak}},
  \bibinfo {author} {\bibfnamefont {D.~W.}\ \bibnamefont {Sumners}}, \bibinfo
  {author} {\bibfnamefont {L.}~\bibnamefont {Tubiana}}, \ and\ \bibinfo
  {author} {\bibfnamefont {C.}~\bibnamefont {Micheletti}},\ }\href@noop {}
  {\bibfield  {journal} {\bibinfo  {journal} {Proc. Nat. Acad. Sci.}\ }\textbf
  {\bibinfo {volume} {106}},\ \bibinfo {pages} {22269} (\bibinfo {year}
  {2009})}\BibitemShut {NoStop}%
\bibitem [{\citenamefont {Rollins}\ \emph {et~al.}(2008)\citenamefont
  {Rollins}, \citenamefont {Petrov},\ and\ \citenamefont {Harvey}}]{Rollins01}%
  \BibitemOpen
  \bibfield  {author} {\bibinfo {author} {\bibfnamefont {G.}~\bibnamefont
  {Rollins}}, \bibinfo {author} {\bibfnamefont {A.}~\bibnamefont {Petrov}}, \
  and\ \bibinfo {author} {\bibfnamefont {S.}~\bibnamefont {Harvey}},\
  }\href@noop {} {\bibfield  {journal} {\bibinfo  {journal} {Biophys. J.}\
  }\textbf {\bibinfo {volume} {94}},\ \bibinfo {pages} {L38} (\bibinfo {year}
  {2008})}\BibitemShut {NoStop}%
\bibitem [{\citenamefont {Petrov}\ \emph {et~al.}(2007)\citenamefont {Petrov},
  \citenamefont {Boz},\ and\ \citenamefont {Harvey}}]{Petrov01}%
  \BibitemOpen
  \bibfield  {author} {\bibinfo {author} {\bibfnamefont {A.}~\bibnamefont
  {Petrov}}, \bibinfo {author} {\bibfnamefont {M.}~\bibnamefont {Boz}}, \ and\
  \bibinfo {author} {\bibfnamefont {S.}~\bibnamefont {Harvey}},\ }\href@noop {}
  {\bibfield  {journal} {\bibinfo  {journal} {J. Struc. Biol.}\ }\textbf
  {\bibinfo {volume} {160}},\ \bibinfo {pages} {241} (\bibinfo {year}
  {2007})}\BibitemShut {NoStop}%
\bibitem [{\citenamefont {LaMarque}\ \emph {et~al.}(2003)\citenamefont
  {LaMarque}, \citenamefont {Le},\ and\ \citenamefont {Harvey}}]{Lamarque01}%
  \BibitemOpen
  \bibfield  {author} {\bibinfo {author} {\bibfnamefont {J.}~\bibnamefont
  {LaMarque}}, \bibinfo {author} {\bibfnamefont {T.}~\bibnamefont {Le}}, \ and\
  \bibinfo {author} {\bibfnamefont {S.}~\bibnamefont {Harvey}},\ }\href@noop {}
  {\bibfield  {journal} {\bibinfo  {journal} {Biopolymers}\ }\textbf {\bibinfo
  {volume} {73}},\ \bibinfo {pages} {348} (\bibinfo {year} {2003})}\BibitemShut
  {NoStop}%
\bibitem [{\citenamefont {Stoop}\ \emph {et~al.}(2011)\citenamefont {Stoop},
  \citenamefont {Najafi}, \citenamefont {Wittel}, \citenamefont {Habibi},\ and\
  \citenamefont {Herrmann}}]{Stoop01}%
  \BibitemOpen
  \bibfield  {author} {\bibinfo {author} {\bibfnamefont {N.}~\bibnamefont
  {Stoop}}, \bibinfo {author} {\bibfnamefont {J.}~\bibnamefont {Najafi}},
  \bibinfo {author} {\bibfnamefont {F.}~\bibnamefont {Wittel}}, \bibinfo
  {author} {\bibfnamefont {M.}~\bibnamefont {Habibi}}, \ and\ \bibinfo {author}
  {\bibfnamefont {H.}~\bibnamefont {Herrmann}},\ }\href@noop {} {\bibfield
  {journal} {\bibinfo  {journal} {Phys. Rev. Lett.}\ }\textbf {\bibinfo
  {volume} {106}},\ \bibinfo {pages} {214102} (\bibinfo {year}
  {2011})}\BibitemShut {NoStop}%
\bibitem [{\citenamefont {Spakowitz}\ and\ \citenamefont
  {Wang}(2005)}]{Spakowitz01}%
  \BibitemOpen
  \bibfield  {author} {\bibinfo {author} {\bibfnamefont {A.}~\bibnamefont
  {Spakowitz}}\ and\ \bibinfo {author} {\bibfnamefont {Z.}~\bibnamefont
  {Wang}},\ }\href@noop {} {\bibfield  {journal} {\bibinfo  {journal} {Biophys.
  J.}\ }\textbf {\bibinfo {volume} {88}},\ \bibinfo {pages} {3912} (\bibinfo
  {year} {2005})}\BibitemShut {NoStop}%
\bibitem [{\citenamefont {Katzav}\ \emph {et~al.}(2006)\citenamefont {Katzav},
  \citenamefont {Adda-Bedia},\ and\ \citenamefont {Boudaoud}}]{Katzav01}%
  \BibitemOpen
  \bibfield  {author} {\bibinfo {author} {\bibfnamefont {E.}~\bibnamefont
  {Katzav}}, \bibinfo {author} {\bibfnamefont {M.}~\bibnamefont {Adda-Bedia}},
  \ and\ \bibinfo {author} {\bibfnamefont {A.}~\bibnamefont {Boudaoud}},\
  }\href@noop {} {\bibfield  {journal} {\bibinfo  {journal} {Proc. Nat. Acad.
  Sci.}\ }\textbf {\bibinfo {volume} {103}},\ \bibinfo {pages} {18900}
  (\bibinfo {year} {2006})}\BibitemShut {NoStop}%
\bibitem [{\citenamefont {Poincare}(1885)}]{Poincare01}%
  \BibitemOpen
  \bibfield  {author} {\bibinfo {author} {\bibfnamefont {H.}~\bibnamefont
  {Poincare}},\ }\href@noop {} {\bibfield  {journal} {\bibinfo  {journal} {J.
  Math. Pures Appl.}\ }\textbf {\bibinfo {volume} {1}},\ \bibinfo {pages} {167}
  (\bibinfo {year} {1885})}\BibitemShut {NoStop}%
\bibitem [{\citenamefont {Hopf}(1926)}]{Hopf01}%
  \BibitemOpen
  \bibfield  {author} {\bibinfo {author} {\bibfnamefont {H.}~\bibnamefont
  {Hopf}},\ }\href@noop {} {\bibfield  {journal} {\bibinfo  {journal} {Math.
  Ann.}\ }\textbf {\bibinfo {volume} {96}},\ \bibinfo {pages} {427} (\bibinfo
  {year} {1926})}\BibitemShut {NoStop}%
\bibitem [{\citenamefont {Nelson}(2002)}]{DNelson01}%
  \BibitemOpen
  \bibfield  {author} {\bibinfo {author} {\bibfnamefont {D.~R.}\ \bibnamefont
  {Nelson}},\ }\href@noop {} {\bibfield  {journal} {\bibinfo  {journal} {Nano
  Letters}\ }\textbf {\bibinfo {volume} {2}},\ \bibinfo {pages} {1125}
  (\bibinfo {year} {2002})}\BibitemShut {NoStop}%
\bibitem [{\citenamefont {Mozaffari}\ \emph {et~al.}(2010)\citenamefont
  {Mozaffari}, \citenamefont {Babadi}, \citenamefont {Fukuda},\ and\
  \citenamefont {Ejtehadi}}]{Mozaffari01}%
  \BibitemOpen
  \bibfield  {author} {\bibinfo {author} {\bibfnamefont {M.}~\bibnamefont
  {Mozaffari}}, \bibinfo {author} {\bibfnamefont {M.}~\bibnamefont {Babadi}},
  \bibinfo {author} {\bibfnamefont {J.}~\bibnamefont {Fukuda}}, \ and\ \bibinfo
  {author} {\bibfnamefont {M.}~\bibnamefont {Ejtehadi}},\ }\href@noop {}
  {\bibfield  {journal} {\bibinfo  {journal} {Soft Matter}\ }\textbf {\bibinfo
  {volume} {7}},\ \bibinfo {pages} {1107} (\bibinfo {year} {2010})}\BibitemShut
  {NoStop}%
\bibitem [{\citenamefont {Angelescu}\ \emph {et~al.}(2008)\citenamefont
  {Angelescu}, \citenamefont {Linse}, \citenamefont {Nguyen},\ and\
  \citenamefont {Bruinsma}}]{Angelescu01}%
  \BibitemOpen
  \bibfield  {author} {\bibinfo {author} {\bibfnamefont {D.~G.}\ \bibnamefont
  {Angelescu}}, \bibinfo {author} {\bibfnamefont {P.}~\bibnamefont {Linse}},
  \bibinfo {author} {\bibfnamefont {T.~T.}\ \bibnamefont {Nguyen}}, \ and\
  \bibinfo {author} {\bibfnamefont {R.~F.}\ \bibnamefont {Bruinsma}},\
  }\href@noop {} {\bibfield  {journal} {\bibinfo  {journal} {Euro. Phys. J. E}\
  }\textbf {\bibinfo {volume} {25}},\ \bibinfo {pages} {323} (\bibinfo {year}
  {2008})}\BibitemShut {NoStop}%
\bibitem [{\citenamefont {Zhang}\ and\ \citenamefont {Chen}(2011)}]{Zhang01}%
  \BibitemOpen
  \bibfield  {author} {\bibinfo {author} {\bibfnamefont {W.~Y.}\ \bibnamefont
  {Zhang}}\ and\ \bibinfo {author} {\bibfnamefont {Z.~Y.}\ \bibnamefont
  {Chen}},\ }\href@noop {} {\bibfield  {journal} {\bibinfo  {journal}
  {Europhys. Lett.}\ }\textbf {\bibinfo {volume} {94}},\ \bibinfo {pages}
  {43001} (\bibinfo {year} {2011})}\BibitemShut {NoStop}%
\bibitem [{\citenamefont {Oskolkov}\ \emph {et~al.}(2011)\citenamefont
  {Oskolkov}, \citenamefont {Linse}, \citenamefont {Potemkin},\ and\
  \citenamefont {Khokhlov}}]{Oskolkov01}%
  \BibitemOpen
  \bibfield  {author} {\bibinfo {author} {\bibfnamefont {N.}~\bibnamefont
  {Oskolkov}}, \bibinfo {author} {\bibfnamefont {P.}~\bibnamefont {Linse}},
  \bibinfo {author} {\bibfnamefont {I.}~\bibnamefont {Potemkin}}, \ and\
  \bibinfo {author} {\bibfnamefont {A.}~\bibnamefont {Khokhlov}},\ }\href@noop
  {} {\bibfield  {journal} {\bibinfo  {journal} {J. Phys. Chem. B}\ }\textbf
  {\bibinfo {volume} {115}},\ \bibinfo {pages} {422} (\bibinfo {year}
  {2011})}\BibitemShut {NoStop}%
\bibitem [{\citenamefont {Fathizadeh}\ \emph {et~al.}(2012)\citenamefont
  {Fathizadeh}, \citenamefont {Eslami-Mossallam},\ and\ \citenamefont
  {Ejtehadi}}]{fathizadeh01}%
  \BibitemOpen
  \bibfield  {author} {\bibinfo {author} {\bibfnamefont {A.}~\bibnamefont
  {Fathizadeh}}, \bibinfo {author} {\bibfnamefont {B.}~\bibnamefont
  {Eslami-Mossallam}}, \ and\ \bibinfo {author} {\bibfnamefont
  {M.}~\bibnamefont {Ejtehadi}},\ }\href@noop {} {\bibfield  {journal}
  {\bibinfo  {journal} {Phys. Rev. E}\ }\textbf {\bibinfo {volume} {86}},\
  \bibinfo {pages} {051907} (\bibinfo {year} {2012})}\BibitemShut {NoStop}%
\bibitem [{\citenamefont {Fathizadeh}\ \emph {et~al.}(2013)\citenamefont
  {Fathizadeh}, \citenamefont {Besya}, \citenamefont {Ejtehadi},\ and\
  \citenamefont {Schiessel}}]{fathizadeh02}%
  \BibitemOpen
  \bibfield  {author} {\bibinfo {author} {\bibfnamefont {A.}~\bibnamefont
  {Fathizadeh}}, \bibinfo {author} {\bibfnamefont {A.~B.}\ \bibnamefont
  {Besya}}, \bibinfo {author} {\bibfnamefont {M.~R.}\ \bibnamefont {Ejtehadi}},
  \ and\ \bibinfo {author} {\bibfnamefont {H.}~\bibnamefont {Schiessel}},\
  }\href@noop {} {\bibfield  {journal} {\bibinfo  {journal} {The European
  Physical Journal E}\ }\textbf {\bibinfo {volume} {36}},\ \bibinfo {pages} {1}
  (\bibinfo {year} {2013})}\BibitemShut {NoStop}%
\bibitem [{\citenamefont {El~Hassan}\ and\ \citenamefont
  {Calladine}(1995)}]{El01}%
  \BibitemOpen
  \bibfield  {author} {\bibinfo {author} {\bibfnamefont {M.}~\bibnamefont
  {El~Hassan}}\ and\ \bibinfo {author} {\bibfnamefont {C.}~\bibnamefont
  {Calladine}},\ }\href@noop {} {\bibfield  {journal} {\bibinfo  {journal} {J.
  Mol. Biol.}\ }\textbf {\bibinfo {volume} {251}},\ \bibinfo {pages} {648}
  (\bibinfo {year} {1995})}\BibitemShut {NoStop}%
\bibitem [{\citenamefont {Olson}\ \emph {et~al.}(2001)\citenamefont {Olson},
  \citenamefont {Bansal}, \citenamefont {Burley}, \citenamefont {Dickerson},\
  and\ \citenamefont {Gerstein}}]{Olson01}%
  \BibitemOpen
  \bibfield  {author} {\bibinfo {author} {\bibfnamefont {W.~K.}\ \bibnamefont
  {Olson}}, \bibinfo {author} {\bibfnamefont {M.}~\bibnamefont {Bansal}},
  \bibinfo {author} {\bibfnamefont {S.~K.}\ \bibnamefont {Burley}}, \bibinfo
  {author} {\bibfnamefont {R.~E.}\ \bibnamefont {Dickerson}}, \ and\ \bibinfo
  {author} {\bibfnamefont {M.}~\bibnamefont {Gerstein}},\ }\href@noop {}
  {\bibfield  {journal} {\bibinfo  {journal} {J. Mol. Biol.}\ }\textbf
  {\bibinfo {volume} {313}},\ \bibinfo {pages} {229} (\bibinfo {year}
  {2001})}\BibitemShut {NoStop}%
\bibitem [{\citenamefont {Becker}\ \emph {et~al.}(2006)\citenamefont {Becker},
  \citenamefont {Wolff},\ and\ \citenamefont {Everaers}}]{Becker01}%
  \BibitemOpen
  \bibfield  {author} {\bibinfo {author} {\bibfnamefont {N.}~\bibnamefont
  {Becker}}, \bibinfo {author} {\bibfnamefont {L.}~\bibnamefont {Wolff}}, \
  and\ \bibinfo {author} {\bibfnamefont {R.}~\bibnamefont {Everaers}},\
  }\href@noop {} {\bibfield  {journal} {\bibinfo  {journal} {Nuc. Acid. Res.}\
  }\textbf {\bibinfo {volume} {34(19)}},\ \bibinfo {pages} {5638} (\bibinfo
  {year} {2006})}\BibitemShut {NoStop}%
\bibitem [{\citenamefont {Becker}\ and\ \citenamefont
  {Everaers}(2007)}]{Becker02}%
  \BibitemOpen
  \bibfield  {author} {\bibinfo {author} {\bibfnamefont {N.}~\bibnamefont
  {Becker}}\ and\ \bibinfo {author} {\bibfnamefont {R.}~\bibnamefont
  {Everaers}},\ }\href@noop {} {\bibfield  {journal} {\bibinfo  {journal}
  {Phys. Rev. E}\ }\textbf {\bibinfo {volume} {76}},\ \bibinfo {pages} {021923}
  (\bibinfo {year} {2007})}\BibitemShut {NoStop}%
\bibitem [{\citenamefont {Lankas}\ \emph {et~al.}(2003)\citenamefont {Lankas},
  \citenamefont {Sponer}, \citenamefont {Langowski},\ and\ \citenamefont
  {Cheatham}}]{Lankas01}%
  \BibitemOpen
  \bibfield  {author} {\bibinfo {author} {\bibfnamefont {F.}~\bibnamefont
  {Lankas}}, \bibinfo {author} {\bibfnamefont {P.}~\bibnamefont {Sponer}},
  \bibinfo {author} {\bibfnamefont {J.}~\bibnamefont {Langowski}}, \ and\
  \bibinfo {author} {\bibfnamefont {T.~E.}\ \bibnamefont {Cheatham}},\
  }\href@noop {} {\bibfield  {journal} {\bibinfo  {journal} {Biophys. J}\
  }\textbf {\bibinfo {volume} {85}},\ \bibinfo {pages} {2872} (\bibinfo {year}
  {2003})}\BibitemShut {NoStop}%
\bibitem [{\citenamefont {Olson}\ \emph {et~al.}(1998)\citenamefont {Olson},
  \citenamefont {Gorin}, \citenamefont {Lu}, \citenamefont {Hock},\ and\
  \citenamefont {Zhurkin}}]{Olson02}%
  \BibitemOpen
  \bibfield  {author} {\bibinfo {author} {\bibfnamefont {W.~K.}\ \bibnamefont
  {Olson}}, \bibinfo {author} {\bibfnamefont {A.}~\bibnamefont {Gorin}},
  \bibinfo {author} {\bibfnamefont {X.}~\bibnamefont {Lu}}, \bibinfo {author}
  {\bibfnamefont {L.}~\bibnamefont {Hock}}, \ and\ \bibinfo {author}
  {\bibfnamefont {V.}~\bibnamefont {Zhurkin}},\ }\href@noop {} {\bibfield
  {journal} {\bibinfo  {journal} {Proc. Nat. Acad. Sci.}\ }\textbf {\bibinfo
  {volume} {95(19)}},\ \bibinfo {pages} {11163} (\bibinfo {year}
  {1998})}\BibitemShut {NoStop}%
\bibitem [{\citenamefont {Everaers}\ and\ \citenamefont
  {Ejtehadi}(2003)}]{Everaers01}%
  \BibitemOpen
  \bibfield  {author} {\bibinfo {author} {\bibfnamefont {R.}~\bibnamefont
  {Everaers}}\ and\ \bibinfo {author} {\bibfnamefont {M.~R.}\ \bibnamefont
  {Ejtehadi}},\ }\href@noop {} {\bibfield  {journal} {\bibinfo  {journal}
  {Phys. Rev. E}\ }\textbf {\bibinfo {volume} {67}},\ \bibinfo {pages} {041710}
  (\bibinfo {year} {2003})}\BibitemShut {NoStop}%
\bibitem [{\citenamefont {Mergell}\ \emph {et~al.}(2003)\citenamefont
  {Mergell}, \citenamefont {Ejtehadi},\ and\ \citenamefont
  {Everaers}}]{Mergell01}%
  \BibitemOpen
  \bibfield  {author} {\bibinfo {author} {\bibfnamefont {B.}~\bibnamefont
  {Mergell}}, \bibinfo {author} {\bibfnamefont {M.~R.}\ \bibnamefont
  {Ejtehadi}}, \ and\ \bibinfo {author} {\bibfnamefont {R.}~\bibnamefont
  {Everaers}},\ }\href@noop {} {\bibfield  {journal} {\bibinfo  {journal}
  {Phys. Rev. E}\ }\textbf {\bibinfo {volume} {68}},\ \bibinfo {pages} {021911}
  (\bibinfo {year} {2003})}\BibitemShut {NoStop}%
\bibitem [{\citenamefont {Gay}\ and\ \citenamefont {Berne}(1981)}]{Gay01}%
  \BibitemOpen
  \bibfield  {author} {\bibinfo {author} {\bibfnamefont {J.~G.}\ \bibnamefont
  {Gay}}\ and\ \bibinfo {author} {\bibfnamefont {B.~J.}\ \bibnamefont
  {Berne}},\ }\href@noop {} {\bibfield  {journal} {\bibinfo  {journal} {J.
  Chem. Phys.}\ }\textbf {\bibinfo {volume} {74}},\ \bibinfo {pages} {3316}
  (\bibinfo {year} {1981})}\BibitemShut {NoStop}%
\bibitem [{\citenamefont {Kamberaj}\ \emph {et~al.}(2005)\citenamefont
  {Kamberaj}, \citenamefont {Low},\ and\ \citenamefont {Neal}}]{Kamberaj01}%
  \BibitemOpen
  \bibfield  {author} {\bibinfo {author} {\bibfnamefont {H.}~\bibnamefont
  {Kamberaj}}, \bibinfo {author} {\bibfnamefont {R.~J.}\ \bibnamefont {Low}}, \
  and\ \bibinfo {author} {\bibfnamefont {M.~P.}\ \bibnamefont {Neal}},\
  }\href@noop {} {\bibfield  {journal} {\bibinfo  {journal} {J. Chem. Phys.}\
  }\textbf {\bibinfo {volume} {122}},\ \bibinfo {pages} {1906216} (\bibinfo
  {year} {2005})}\BibitemShut {NoStop}%
\bibitem [{\citenamefont {Babadi}\ \emph {et~al.}(2006)\citenamefont {Babadi},
  \citenamefont {Ejtehadi},\ and\ \citenamefont {Everaers}}]{Babadi01}%
  \BibitemOpen
  \bibfield  {author} {\bibinfo {author} {\bibfnamefont {M.}~\bibnamefont
  {Babadi}}, \bibinfo {author} {\bibfnamefont {M.~R.}\ \bibnamefont
  {Ejtehadi}}, \ and\ \bibinfo {author} {\bibfnamefont {R.}~\bibnamefont
  {Everaers}},\ }\href@noop {} {\bibfield  {journal} {\bibinfo  {journal} {J.
  Comp. Phys.}\ }\textbf {\bibinfo {volume} {219}},\ \bibinfo {pages} {770}
  (\bibinfo {year} {2006})}\BibitemShut {NoStop}%
\bibitem [{\citenamefont {Smith}\ \emph {et~al.}(2001)\citenamefont {Smith},
  \citenamefont {Tans}, \citenamefont {Smith}, \citenamefont {Grimes},
  \citenamefont {Anderson},\ and\ \citenamefont {Bustamante}}]{Smith02}%
  \BibitemOpen
  \bibfield  {author} {\bibinfo {author} {\bibfnamefont {D.}~\bibnamefont
  {Smith}}, \bibinfo {author} {\bibfnamefont {S.}~\bibnamefont {Tans}},
  \bibinfo {author} {\bibfnamefont {S.}~\bibnamefont {Smith}}, \bibinfo
  {author} {\bibfnamefont {S.}~\bibnamefont {Grimes}}, \bibinfo {author}
  {\bibfnamefont {D.}~\bibnamefont {Anderson}}, \ and\ \bibinfo {author}
  {\bibfnamefont {C.}~\bibnamefont {Bustamante}},\ }\href@noop {} {\bibfield
  {journal} {\bibinfo  {journal} {Nature}\ }\textbf {\bibinfo {volume} {413}},\
  \bibinfo {pages} {748} (\bibinfo {year} {2001})}\BibitemShut {NoStop}%
\bibitem [{\citenamefont {Fuller}\ \emph
  {et~al.}(2007{\natexlab{a}})\citenamefont {Fuller}, \citenamefont {Raymer},
  \citenamefont {Rickgauer}, \citenamefont {Robertson}, \citenamefont
  {Catalano}, \citenamefont {Anderson}, \citenamefont {Grimes},\ and\
  \citenamefont {Smith}}]{Fuller01}%
  \BibitemOpen
  \bibfield  {author} {\bibinfo {author} {\bibfnamefont {D.}~\bibnamefont
  {Fuller}}, \bibinfo {author} {\bibfnamefont {D.}~\bibnamefont {Raymer}},
  \bibinfo {author} {\bibfnamefont {J.}~\bibnamefont {Rickgauer}}, \bibinfo
  {author} {\bibfnamefont {R.}~\bibnamefont {Robertson}}, \bibinfo {author}
  {\bibfnamefont {C.}~\bibnamefont {Catalano}}, \bibinfo {author}
  {\bibfnamefont {D.}~\bibnamefont {Anderson}}, \bibinfo {author}
  {\bibfnamefont {S.}~\bibnamefont {Grimes}}, \ and\ \bibinfo {author}
  {\bibfnamefont {D.}~\bibnamefont {Smith}},\ }\href@noop {} {\bibfield
  {journal} {\bibinfo  {journal} {J. Mol. Biol.}\ }\textbf {\bibinfo {volume}
  {373}},\ \bibinfo {pages} {1113} (\bibinfo {year}
  {2007}{\natexlab{a}})}\BibitemShut {NoStop}%
\bibitem [{\citenamefont {Fuller}\ \emph
  {et~al.}(2007{\natexlab{b}})\citenamefont {Fuller}, \citenamefont {Raymer},
  \citenamefont {Kottadiel}, \citenamefont {Rao},\ and\ \citenamefont
  {Smith}}]{Fuller02}%
  \BibitemOpen
  \bibfield  {author} {\bibinfo {author} {\bibfnamefont {D.}~\bibnamefont
  {Fuller}}, \bibinfo {author} {\bibfnamefont {D.}~\bibnamefont {Raymer}},
  \bibinfo {author} {\bibfnamefont {V.}~\bibnamefont {Kottadiel}}, \bibinfo
  {author} {\bibfnamefont {V.}~\bibnamefont {Rao}}, \ and\ \bibinfo {author}
  {\bibfnamefont {D.}~\bibnamefont {Smith}},\ }\href@noop {} {\bibfield
  {journal} {\bibinfo  {journal} {Proc. Nat. Acad. Sci.}\ }\textbf {\bibinfo
  {volume} {104}},\ \bibinfo {pages} {16868} (\bibinfo {year}
  {2007}{\natexlab{b}})}\BibitemShut {NoStop}%

\bibitem [{\citenamefont {Kl{\'e}man}\ and\ \citenamefont
  {Laverntovich}(2002)}]{Kleman01}%
  \BibitemOpen
  \bibfield  {author} {\bibinfo {author} {\bibfnamefont {M.}~\bibnamefont
  {Kl{\'e}man}}\ and\ \bibinfo {author} {\bibfnamefont {O.}~\bibnamefont
  {Laverntovich}},\ }\href@noop {} {\emph {\bibinfo {title} {Soft Matter
  Physics: an Introduction}}}\ (\bibinfo  {publisher} {Springer},\ \bibinfo
  {year} {2002})\BibitemShut {NoStop}%
\bibitem [{\citenamefont {Bloomfield}(1997)}]{Bloomfield01}%
  \BibitemOpen
  \bibfield  {author} {\bibinfo {author} {\bibfnamefont {V.}~\bibnamefont
  {Bloomfield}},\ }\href@noop {} {\bibfield  {journal} {\bibinfo  {journal}
  {Biopolymers}\ }\textbf {\bibinfo {volume} {44}},\ \bibinfo {pages} {269}
  (\bibinfo {year} {1997})}\BibitemShut {NoStop}%
\bibitem [{\citenamefont {Kindt}\ \emph {et~al.}(2001)\citenamefont {Kindt},
  \citenamefont {Tzlil}, \citenamefont {Ben-Shaul},\ and\ \citenamefont
  {Gelbart}}]{Kindt01}%
  \BibitemOpen
  \bibfield  {author} {\bibinfo {author} {\bibfnamefont {J.}~\bibnamefont
  {Kindt}}, \bibinfo {author} {\bibfnamefont {S.}~\bibnamefont {Tzlil}},
  \bibinfo {author} {\bibfnamefont {A.}~\bibnamefont {Ben-Shaul}}, \ and\
  \bibinfo {author} {\bibfnamefont {W.}~\bibnamefont {Gelbart}},\ }\href@noop
  {} {\bibfield  {journal} {\bibinfo  {journal} {Proc. Nat. Acad. Sci.}\
  }\textbf {\bibinfo {volume} {98}},\ \bibinfo {pages} {13671} (\bibinfo {year}
  {2001})}\BibitemShut {NoStop}%
\bibitem [{\citenamefont {Sottas}\ \emph {et~al.}(1999)\citenamefont {Sottas},
  \citenamefont {Larquet}, \citenamefont {Stasiak},\ and\ \citenamefont
  {Dubochet}}]{Sottas01}%
  \BibitemOpen
  \bibfield  {author} {\bibinfo {author} {\bibfnamefont {P.}~\bibnamefont
  {Sottas}}, \bibinfo {author} {\bibfnamefont {E.}~\bibnamefont {Larquet}},
  \bibinfo {author} {\bibfnamefont {A.}~\bibnamefont {Stasiak}}, \ and\
  \bibinfo {author} {\bibfnamefont {J.}~\bibnamefont {Dubochet}},\ }\href@noop
  {} {\bibfield  {journal} {\bibinfo  {journal} {Biophys. J.}\ }\textbf
  {\bibinfo {volume} {77}},\ \bibinfo {pages} {1858} (\bibinfo {year}
  {1999})}\BibitemShut {NoStop}%
\bibitem [{\citenamefont {Angelescu}\ and\ \citenamefont
  {Linse}(2008)}]{angelescu03}%
  \BibitemOpen
  \bibfield  {author} {\bibinfo {author} {\bibfnamefont {D.}~\bibnamefont
  {Angelescu}}\ and\ \bibinfo {author} {\bibfnamefont {P.}~\bibnamefont
  {Linse}},\ }\href@noop {} {\bibfield  {journal} {\bibinfo  {journal} {Soft
  Matter}\ }\textbf {\bibinfo {volume} {4}},\ \bibinfo {pages} {1981} (\bibinfo
  {year} {2008})}\BibitemShut {NoStop}%
  

\end{thebibliography}
\end{document}